\documentclass[useAMS,usenatbib,usegraphicx]{mn2e}

\usepackage{times}

\usepackage{color}
\usepackage{soul}
\usepackage{hyperref}

\usepackage[letterpaper,margin=0.8in]{geometry}
\usepackage{geometry}

\title[Correlated variability and high energy emission in blazars]{Time-correlation between the radio and gamma-ray activity in blazars and the production site of the gamma-ray emission}

\author[W. Max-Moerbeck et al.]{
W.~Max-Moerbeck$^{1,2}$\thanks{E-mail: wmax@nrao.edu},  
T.~Hovatta$^{1,3}$,
J.~L.~Richards$^{4}$,
O.~G.~King$^{1}$,
T.~J.~Pearson$^{1}$, \newauthor
A.~C.~S.~Readhead$^{1}$,
R.~Reeves$^{1,9}$,
M.~C.~Shepherd$^{1}$,
M.~A.~Stevenson$^{1}$,
E.~Angelakis$^{5}$, \newauthor
L.~Fuhrmann$^{5}$,
K.~J.~B.~Grainge$^{7}$,
V.~Pavlidou$^{1,5,6}$,
R.~W.~Romani$^{8}$,
J.~A.~Zensus$^{5}$\\
$^{1}$Cahill Center for Astronomy and Astrophysics, California Institute of Technology, Pasadena, CA 91125, USA\\
$^{2}$National Radio Astronomy Observatory (NRAO), P.O. Box 0, Socorro, NM 87801, USA\\
$^{3}$Aalto University Mets\"ahovi Radio Observatory, Mets\"ahovintie 114, 02540 Kylm\"al\"a, Finland\\
$^{4}$Department of Physics, Purdue University, West Lafayette, IN 47907, USA\\
$^{5}$Max-Planck-Institut f\"ur Radioastronomie, Auf dem H\"ugel 69, 53121 Bonn, Germany\\
$^{6}$Department of Physics, University of Crete / Foundation for Research and Technology - Hellas, Heraklion 71003, Greece\\
$^{7}$Jodrell Bank Centre for Astrophysics, School of Physics and Astronomy, The University of Manchester, M13 9PL\\
$^{8}$W. W. Hansen Experimental Physics Laboratory, Kavli Institute for Particle Astrophysics and Cosmology, Department of Physics and\\
SLAC National Accelerator Laboratory, Stanford University, Stanford, CA 94305, USA\\
$^{9}$Departamento de Astronom\'ia, Universidad de Concepci\'on, Casilla 160-C, Concepci\'on, Chile
}

\begin{document}

\date{Accepted 2014 August 26. Received 2014 August 21; in original form 2014 June 20}

\pagerange{\pageref{firstpage}--\pageref{lastpage}} \pubyear{0000}

\maketitle

\label{firstpage}

\begin{abstract}
In order to determine the location of the gamma-ray emission site in blazars, we investigate the time-domain relationship between their radio and gamma-ray emission. Light-curves for the brightest detected blazars from the first 3 years of the mission of the \emph{Fermi Gamma-ray Space Telescope} are cross-correlated with 4 years of 15\,GHz observations from the OVRO 40-m monitoring program. The large sample and long light-curve duration enable us to carry out a statistically robust analysis of the significance of the cross-correlations, which is investigated using Monte Carlo simulations including the uneven sampling and noise properties of the light-curves. Modeling the light-curves as red noise processes with power-law power spectral densities, we find that only one of 41 sources with high quality data in both bands shows correlations with significance larger than 3$\sigma$ (AO\,0235+164), with only two more larger than even 2.25$\sigma$ (PKS\,1502+106 and B2\,2308+34). Additionally, we find correlated variability in Mrk\,421 when including a strong flare that occurred in July-September 2012. These results demonstrate very clearly the difficulty of measuring statistically robust multiwavelength correlations and the care needed when comparing light-curves even when many years of data are used.  This should be a caution. In all four sources the radio variations lag the gamma-ray variations, suggesting that the gamma-ray emission originates upstream of the radio emission. Continuous simultaneous monitoring over a longer time period is required to obtain high significance levels in cross-correlations between gamma-ray and radio variability in most blazars.
\end{abstract}


\section{Introduction}

Blazars are active galactic nuclei with jets closely aligned to the line of sight \citep[e.g.,][]{blandford_1979}. They are the most numerous class of sources detected in the\,GeV band by the Large Area Telescope (LAT) on the \emph{Fermi Gamma-ray Space Telescope} \citep{2LAC_2011}. Blazars have double-peaked broad-band spectral energy distributions and show strong variability from radio to gamma-rays \citep[e.g.,][]{vonmontingny+1995}. It is accepted that the low-energy emission is produced by synchrotron radiation from electrons within the jet, while the high-energy gamma-ray emission is produced by inverse-Compton scattering of a soft photon field by the same electrons \citep[e.g.,][]{jones+1974, dermer+1993, sikora+1994, blazejowski+2000} or by hadronic processes \citep[e.g.,][]{mannheim_biermann1992}. That a common mechanism regulates the luminosity at high and low energies is demonstrated by the correlation between the mean radio flux density and mean gamma-ray flux \citep[][]{kovalev_2009, mahony_2010, nieppola+2011}. \citet{ackermann_2011} and \citet{pavlidou+2012} showed that this correlation is not an effect of distance modulation of the fluxes.

The location of the gamma-ray emission site in blazars is not yet known. Gamma rays may be produced, for example, in the radio-emitting  regions \citep[e.g.,][]{jorstad_2001}, or much closer to the central engine \citep[e.g.,][]{blandford_and_levinson_1995}. Radio observations with milli-arcsecond resolution have resolved the radio-emitting regions and measured outflow velocities, but at high energies the angular resolution is insufficient and we must infer the size and location of the emission regions from flux variations. If gamma-ray and radio emission are triggered by shocks propagating along a relativistic jet, the time delay between flares in the two bands depends on their separation. Several studies have found time-lagged correlation between these two energy bands, but without a large sample with well-sampled light-curves it is difficult to assess the significance of the correlations \citep[e.g.,][]{marscher_bllac_2008, abdo_3c279_2010, agudo_oj_2011, agudo_ao_2011}. In a statistical study of 183 bright \textit{Fermi}-detected sources, \citet{pushkarev_2010} found that, on average, radio flares occur later than gamma-ray flares. A more recent investigation using multiple radio frequencies and longer light-curves \citep{fuhrmann+2014} also found correlated radio and gamma-ray variability with a frequency dependent radio lag.

In comparing multiwavelength light-curves of individual blazars over short time periods claims are often made for correlations but the actual significance is rarely computed. To remedy this situation and search for the existence of significant correlations and their physical origin we have undertaken a long-term radio monitoring campaign of a large number of blazars. We apply robust statistical methods to estimate the significance of correlations and find that most of the blazars in our sample only show correlations below 2.25$\sigma$.  Only three out of 41 objects show correlations above a 2.25$\sigma$ level where we expect to find one random uncorrelated source to appear, with only one above the 3$\sigma$ level of significance. Thus, it is clear that establishing a statistically significant cross-correlation is more difficult than is generally assumed. We also provide a tentative interpretation for the origin of the time lag and the location of the gamma-ray emission site.

\section{Observations}
\label{observations}

Through our Owens Valley Radio Observatory (OVRO) 40-m program, twice per week we observe all sources in the Candidate Gamma Ray Blazar Survey \citep[CGRaBS,][]{healey+2008} and the blazars detected in the \textit{Fermi}-LAT AGN catalogs \citep[][]{1LAC_2010, 2LAC_2011} north of declination $-20$\degr~at 15\,GHz. This sample has a total of 1,593 sources, of which 685 have gamma-ray detections, with 454 and 634 in the first and second \textit{Fermi}-LAT AGN catalogs respectively. 

Radio observations from 1 January 2008 to 26 February 2012 are included in this study. The radio flux density measurements have a thermal noise floor of $\sim5$ mJy with an additional 2\% contribution from pointing errors. The flux density scale is determined from regular observations of 3C 286 assuming the \citet[][]{baars+1977} value of 3.44 Jy at 15.0 GHz, giving a 5\% overall scale accuracy. A detailed discussion of the observing strategy and calibration procedures can be found in \citet[][]{richards+2011}. The radio light-curves have different characteristics, with a mean and standard deviation for length $1178 \pm 441$ days, number of data points $195 \pm 88$, and average sampling $6.4 \pm 1.4$ days. The light-curves of the cases discussed in this paper are shown in Figure \ref{xcorr_combined}. The monitoring program is ongoing and all the light-curves are made public on the program website\footnote{\url{http://astro.caltech.edu/ovroblazars/}}.

The LAT is a pair-conversion gamma-ray telescope, sensitive to photon energies from about 20 MeV up to $>$300 GeV, that observes the whole sky once every three hours \citep[][]{atwood+2009}. \emph{Fermi}-LAT light-curves with 7\,d time bins from 4 August 2008 through 12 August 2011 were produced for 86 sources detected in at least 75\% of monthly time bins \citep[][]{2FGL_2012}. We use an unbinned likelihood analysis, with source spectral models and positions from \citet[][]{2LAC_2011}. We froze the sources spectral parameters (including the target) and let only the flux vary in sources within 10\degr~of the target. We use \emph{Fermi}-LAT \texttt{ScienceTools-v9r23p1} with P7\_V6 source event selection and instrument response functions, diffuse models \texttt{gal\_2yearp7v6\_v0.fits} and \texttt{iso\_p7v6source.txt}, only photons with zenith angle $<$100\degr~and other standard data cuts and filters \citep[e.g.,][]{abdo_bllac_2011}\footnote{Science Tools, LAT data, and diffuse emission models are available from the \textit{Fermi} Science Support Center, \url{http://fermi.gsfc.nasa.gov/ssc}}. We use a region of interest of 10\degr~radius and a source region of 15\degr~radius. Photon integral fluxes from 100 MeV to 200\,GeV are reported when the test statistic\footnote{The test statistic is a measure of detection significance, defined as TS$=2\Delta\log({\rm likelihood})$ between models with and without the source \citep[][]{mattox+1996}.} ${\rm TS}\ge4$, and $2\sigma$ upper limits when ${\rm TS}<4$ ($\sim 30$\% of the data).

\section{Time lags and their significance}
\label{time-lags-and-significance}

The radio light-curves are sampled unevenly due to weather and other problems. The gamma-ray light-curves are weekly averages, but some measurements are upper limits ($\sim 30$\% of the data) that are ignored in this analysis. We tested the possible effect of ignoring upper limits by using the best flux estimate independent of TS and the upper limit itself as a flux, obtaining comparable results in all cases, thus showing that it is safe to ignore upper limits for this sample of bright sources. The cross-correlation is measured using the discrete cross-correlation function \citep[DCF,][]{edelson_1988}, with local normalization \citep[][]{welsh_1999}, also known as local cross-correlation function (LCCF). We find that the LCCF results in a greater detection efficiency for known correlations injected in simulated data. We estimate the cross-correlation significance with Monte Carlo simulations that assume a simple power-law power spectral density model for the light-curves (${\rm PSD} \propto1/f^{\beta}$), motivated by previous work \citep[e.g.,][]{hufnagel_1992, edelson_1995, uttley_2003, arevalo_2008, chatterjee_2008, abdo_variability_2010}. We simulate a large number of independent, uncorrelated light-curve pairs that replicate the sampling, measurement error distribution, and statistical properties of the observations, using  the method of \citet[][]{timmer+1995}. From the distribution of cross-correlations at each time lag we estimate the chance probability of obtaining a given correlation value. The method is described in detail by \citet[][]{max-moerbeck+2014}

For 13 sources where a PSD fit is possible in both bands, we use the best-fitting power-law index values; for the others, we use population-average values as described below. We characterize the PSDs using a modified implementation of \citet[][]{uttley+2002} that uses sampling window functions to reduce red-noise leakage. The effects of uneven sampling are incorporated by comparing the observed PSD to those derived from simulated light-curves. We compute the PSD from the data and obtain a mean PSD with scatter from simulated light-curves for several values of the power-law index. The best fit is found by comparing the PSD from the data with the simulated ones using a $\chi^{2}$ test. We find good constraints for the radio PSD power-law index for 43 sources (Table \ref{xcorr_sig_table}). The distribution of indices is clustered around 2.3, with a typical error of 0.4, and is consistent with a single value equal to the sample mean of $2.3\pm0.1$. We adopt a value of $\beta_{\rm radio}=2.3$ for sources with no fitted radio PSD. In the gamma-ray band the PSD power-law index is constrained for 29 sources. The distribution has peaks at about 0.5 and 1.6. The peak at 1.6 is consistent with results for the brightest sources from \citet[][]{abdo_variability_2010} ($1.4\pm0.1$ for FSRQs and $1.7\pm0.3$ for BL Lacs) but steeper than found in \citet[][]{2LAC_2011} (about 1.15 for the average PSD of the brightest blazars). For sources with no gamma-ray PSD fit, we assume $\beta_{\rm \gamma}=1.6$ which gives conservative estimates of the cross-correlation significance.

\section{Results of the cross-correlation significance}

We estimated the cross-correlation between the radio and gamma-ray light-curves and its significance for 41 of the 86 sources. 23 are excluded for being non-variable at the 3$\sigma$ level  (a $\chi^2$ test of the null hypothesis of constant flux shows that the observed variations are consistent with observational noise). We also exclude ``noisy'' light-curves where more than $1/3$ of the variance comes from observational noise. We also exclude light-curves consistent with a linear trend in the overlapping section; for such sources, longer light-curves are needed to probe the relevant time scales. These two restrictions eliminate 22 more objects.

To include the effects of red-noise leakage and aliasing we simulate 10-year light-curves with a 1\,d time resolution. The cross-correlation is estimated for independent bins of 10\,d. In each case, we simulate 20,000 independent light-curve pairs using the appropriate PSD (Section \ref{time-lags-and-significance} and Table \ref{xcorr_sig_table}). To eliminate spurious correlations we restrict the time lag search interval to $\pm0.5$ times the length of the shortest light-curve. For each source the position and significance of the most significant cross-correlation peak is given in Table \ref{xcorr_sig_table}. The peak position uncertainty is estimated by ``flux randomization'' and ``random subset selection'' \citep[][]{peterson+1998}. The error on the significance is determined using a bootstrap method \citep[][]{max-moerbeck+2014}. We set the significance threshold at 97.56\% ($2.25\sigma$), at which we expect to have one object with a chance high correlation.

\begin{table*}
\caption{Cross-correlation significance results \label{xcorr_sig_table} (see complete table at the end of the document. The journal version has the table in landscape mode.)}
\end{table*}

\begin{figure*}
\includegraphics[angle=0,width=8.5cm, trim=30 0 30 5]{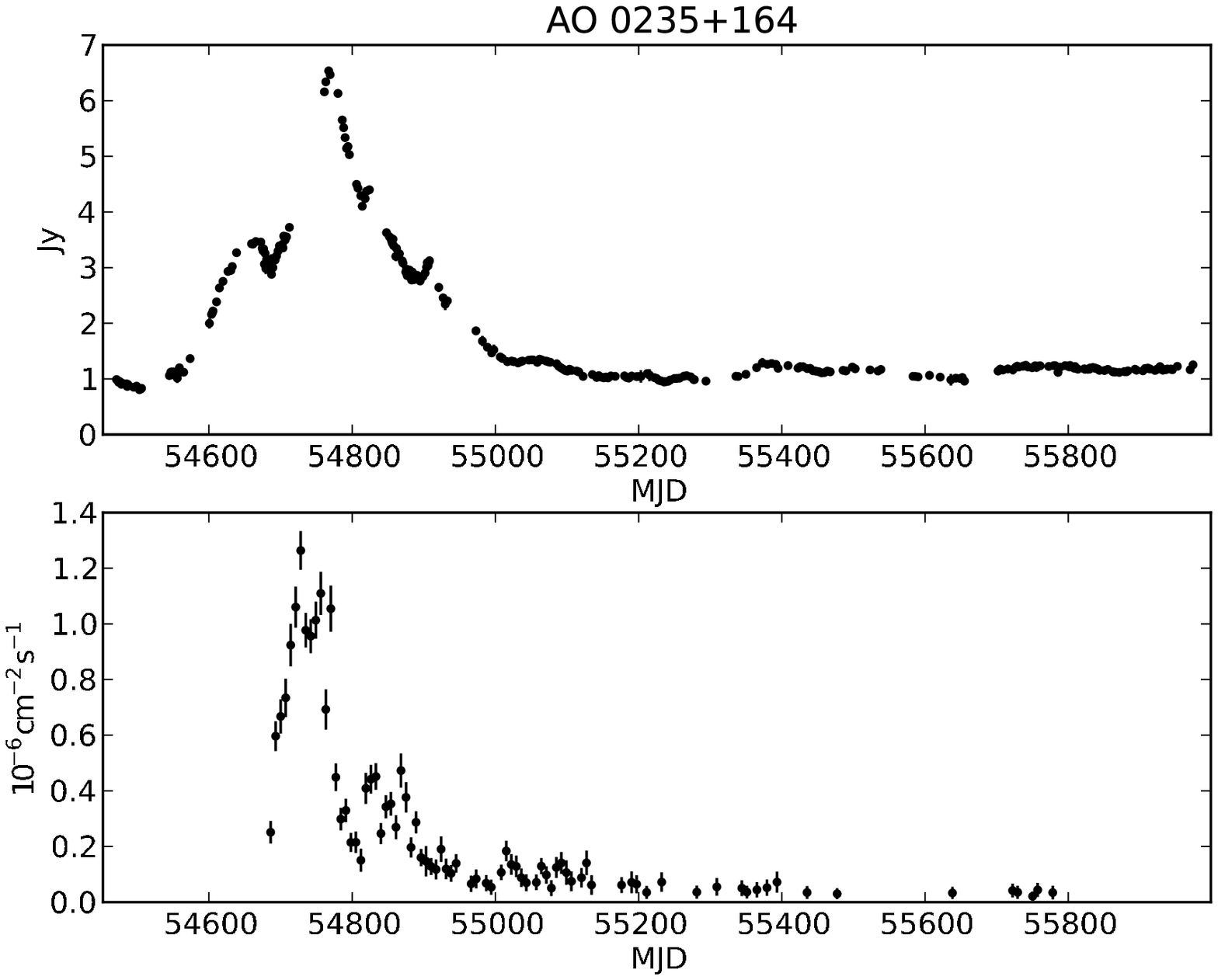}
\includegraphics[angle=0,width=8.5cm, trim=30 0 30 5]{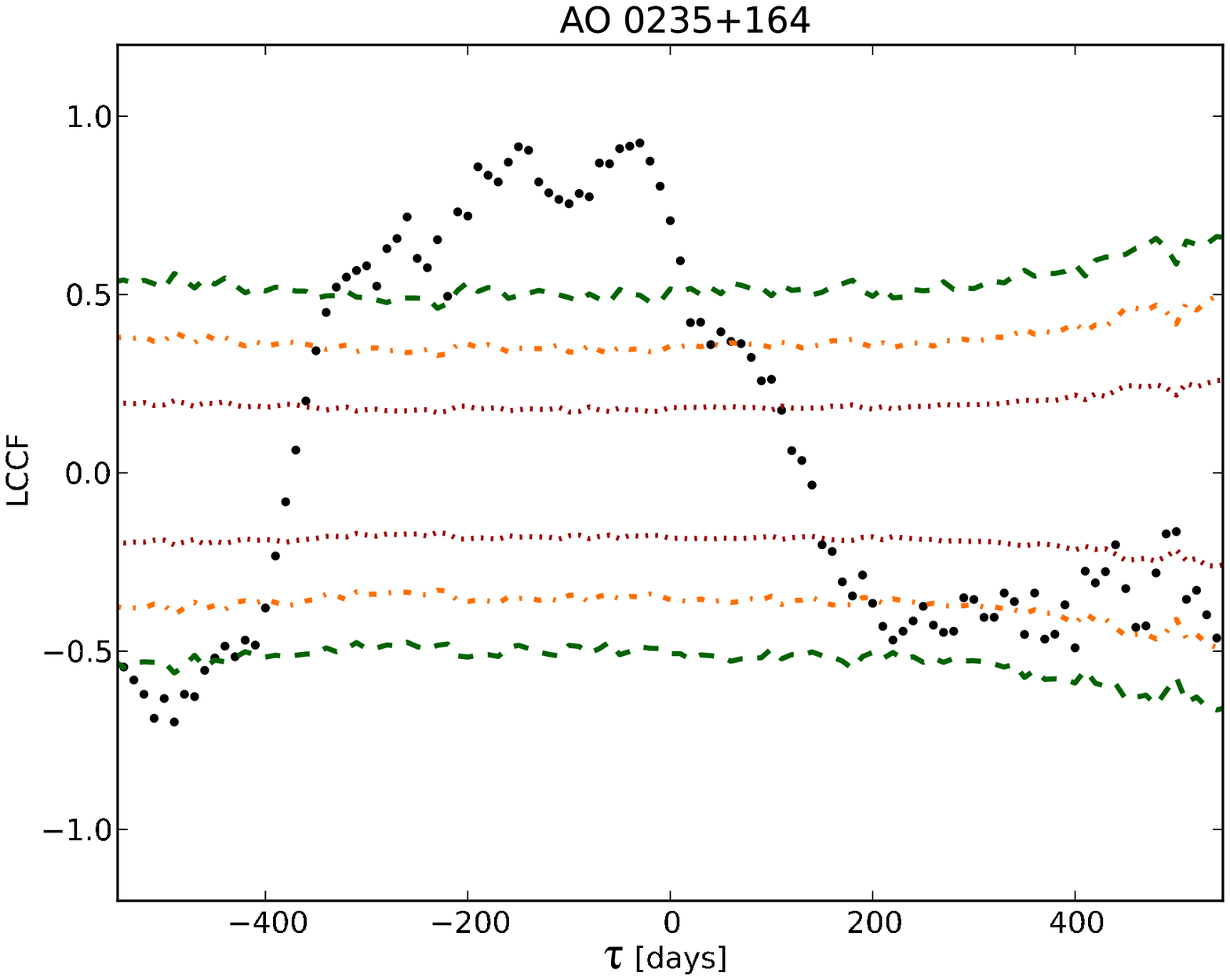}
\includegraphics[angle=0,width=8.5cm, trim=30 0 30 5]{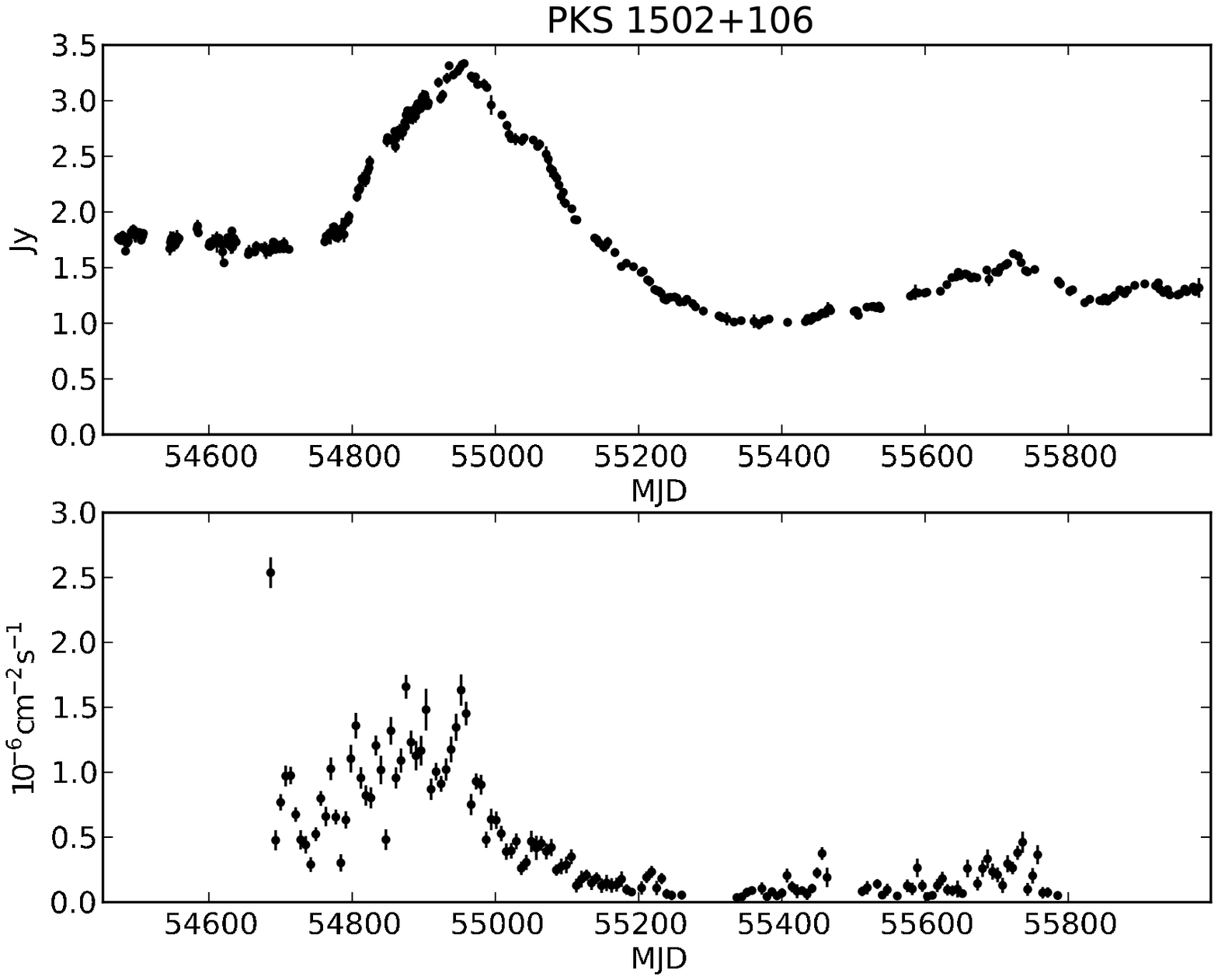}
\includegraphics[angle=0,width=8.5cm, trim=30 0 30 5]{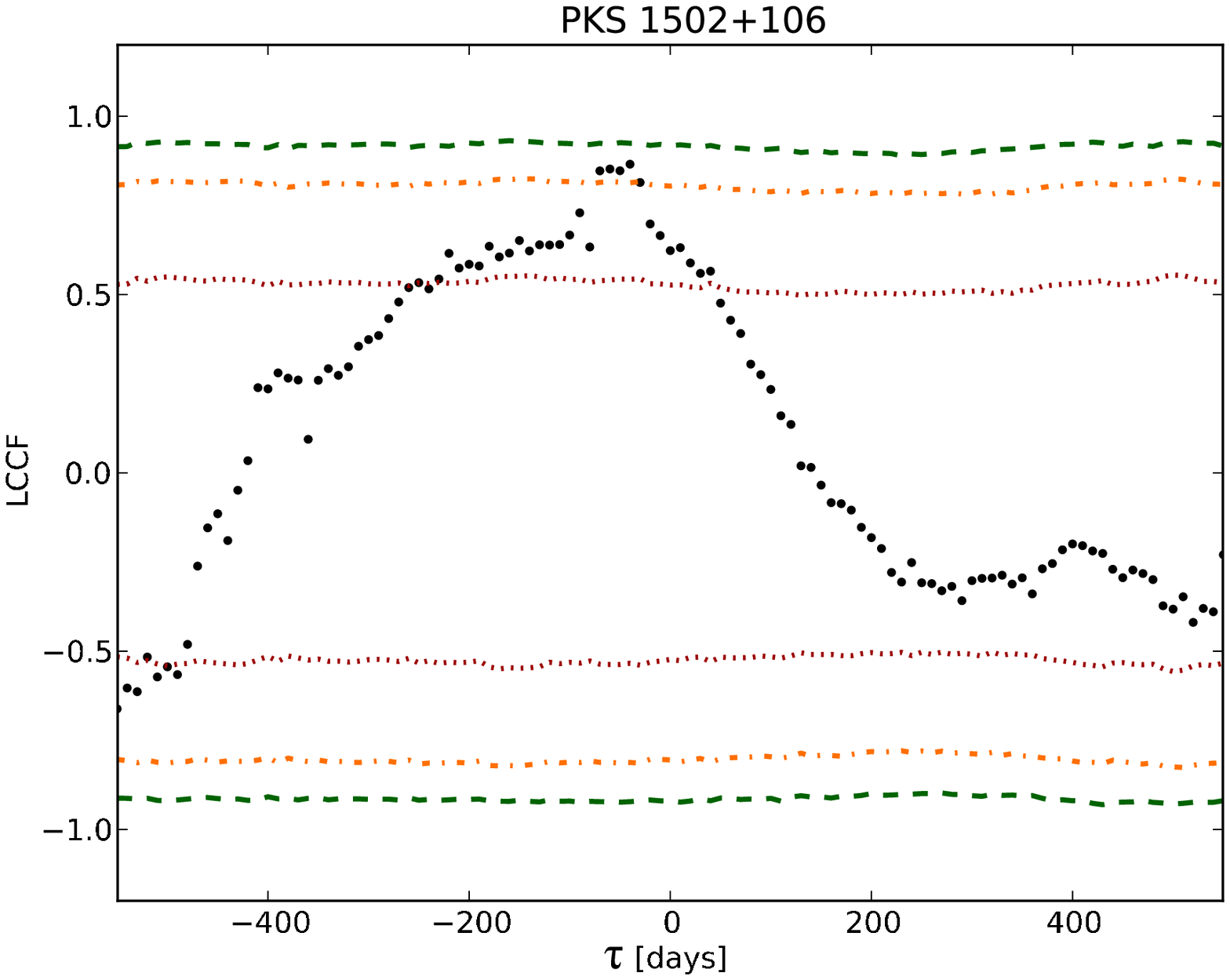}
\includegraphics[angle=0,width=8.5cm, trim=30 0 30 5]{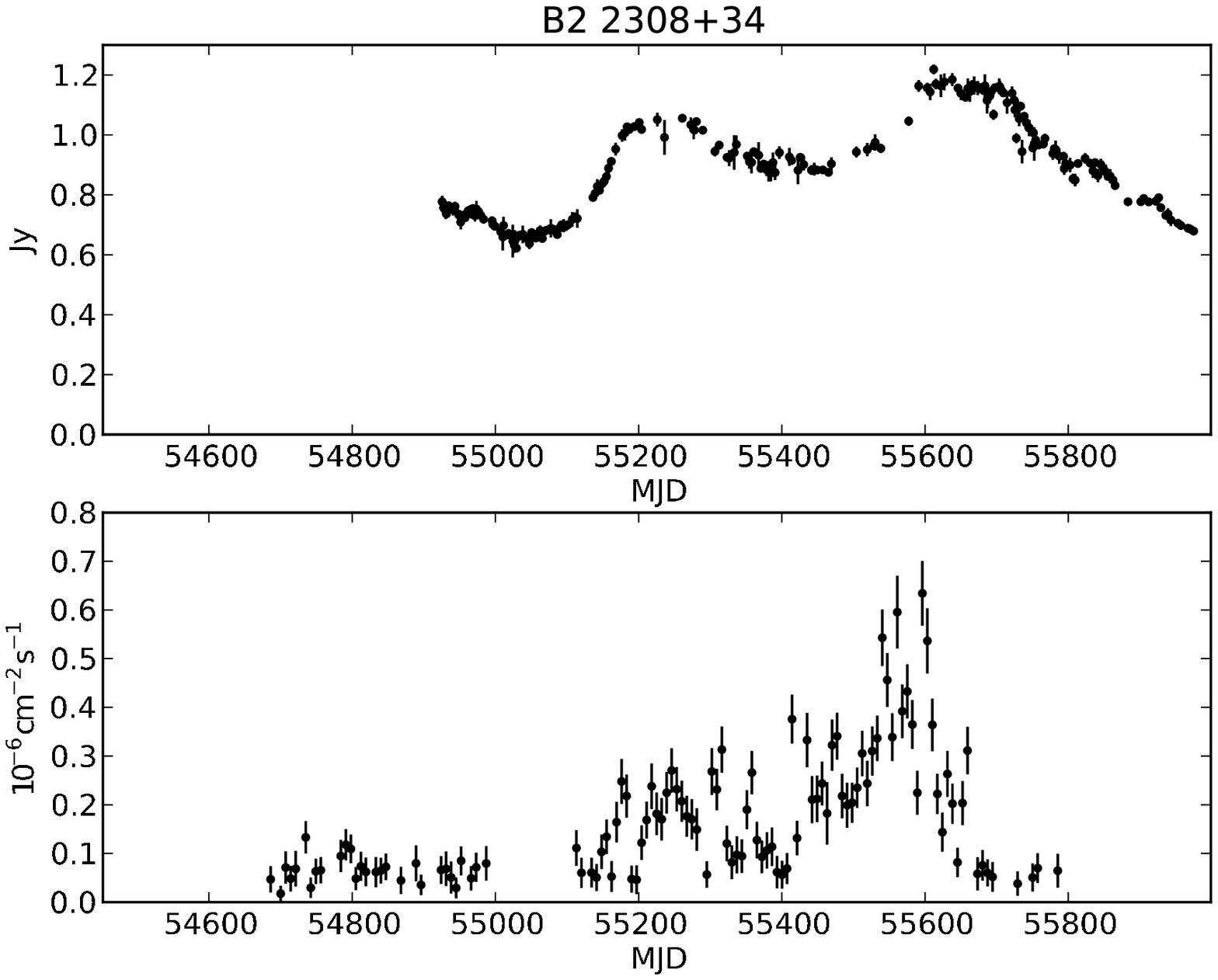}
\includegraphics[angle=0,width=8.5cm, trim=30 0 30 5]{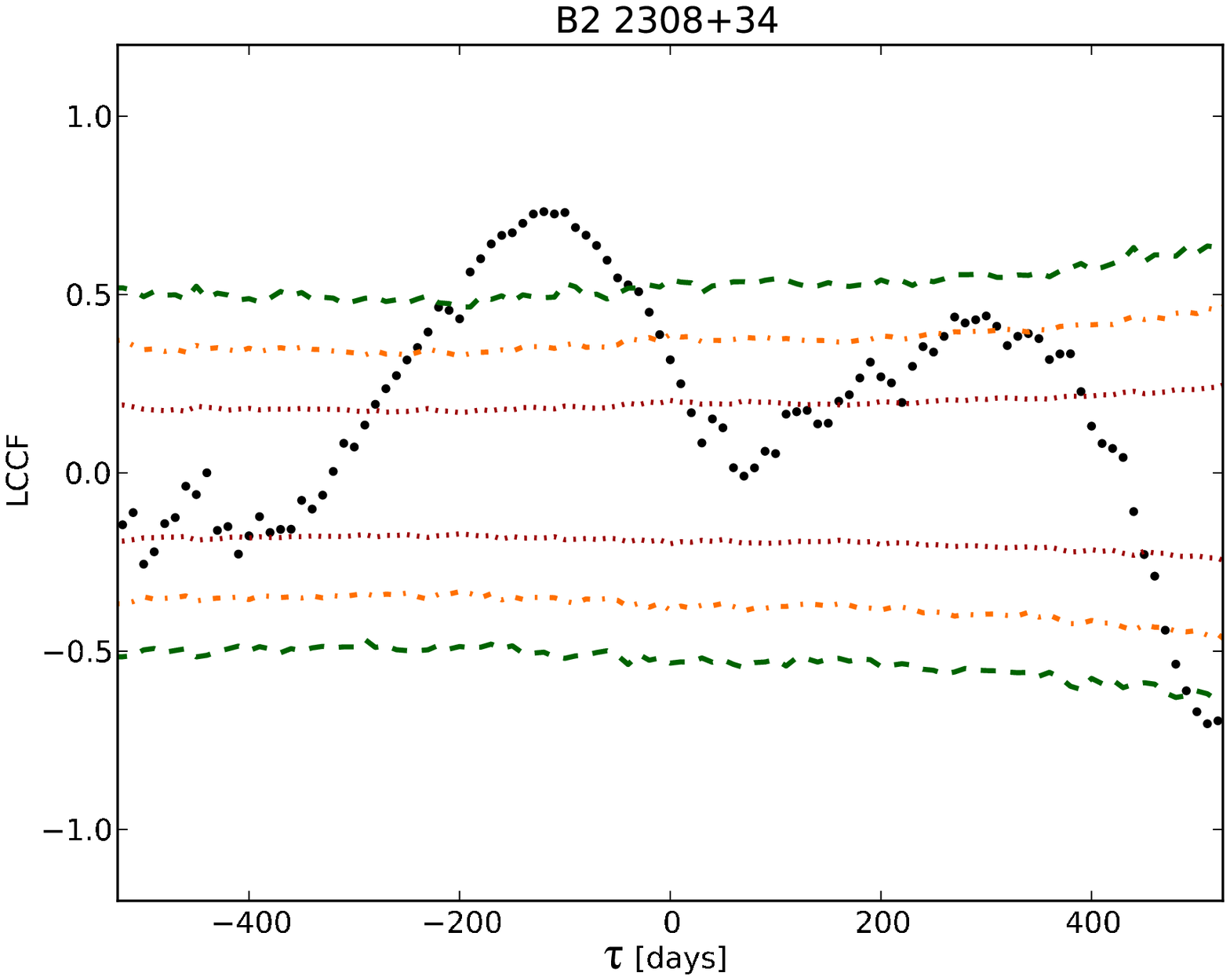}
\caption{Light-curves (left) and cross-correlation (right) for sources with significant cross-correlation. Contours indicate the cross-correlations significances (red dotted line: $1\sigma$; orange dash-dot line: $2\sigma$; green dashed line: $3\sigma$). The most significant peak for AO~0235+164 is at $-150\pm8\,$d with 99.99\% significance, for PKS\,1502+106 is at $-40\pm13\,$d with 98.09\% significance for the best fit PSD model and 97.54\% for the lower limit, and for B2 2308+34 is at $-120\pm14\,$d with 99.99\% significance for the best fit PSD model and 99.33\% for the lower limit. The significance lower limit for PKS\,1502+106 is above the 97.56\% threshold within the error (see Table \ref{xcorr_sig_table}).}
\label{xcorr_combined}
\end{figure*}

At this threshold, three of our 41 sources show interesting levels of correlation: AO\,0235+164, $\tau=-150\pm8\,$d with 99.99\% significance (the only case with significance $\ge3\sigma$); PKS\,1502+106, $\tau=-40\pm13\,$d with 97.54\% significance\footnote{This is consistent with the threshold of 97.56\% when the 0.13\% uncertainty is considered as shown in Table \ref{xcorr_sig_table}}; and B2\,2308+34, $\tau=-120\pm14\,$d with 99.33\% significance. The results are presented in Figure~\ref{xcorr_combined}, where a negative lag indicates that radio variations occur after gamma-ray variations.

Significant correlated variability has been reported by \citet[][]{agudo_ao_2011} for AO\,0235+164, with a delay of about $-30\,$d using radio data up to MJD~55000. With our longer light-curves, we find a significant correlation at a delay of $-150\,$d, although the cross-correlation peak is broad and there is a second peak of comparable amplitude and significance at $-30\,$d. This adds a large uncertainty when considered in the estimation of the location of the gamma-ray emission site because our current data cannot discriminate between these two peaks. No significant cross-correlations have been previously reported for PKS\,1502+106 or B2\,2304+34.

\section{The case of Mrk\,421}

A major radio flare was observed from Mrk\,421 on 21 September 2012, when its 15\,GHz flux density reached $1.11\pm0.03$~Jy, approximately 2.5 times its previous median value \citep[][]{ovro_mrk421_atel_2012}. On 16 July 2012, the source was detected at its highest level to date by \emph{Fermi}-LAT. Its integrated photon flux for ${\rm E}>100$\,MeV was $(1.4\pm0.2)\times10^{-6}$\,ph\,cm$^{-2}$\,s$^{-1}$, a factor of 8 greater than the average in the second \emph{Fermi}-LAT catalog \citep[][]{fermi_mrk421_atel_2012}.

\begin{figure*}
\includegraphics[angle=0,width=8.5cm, trim=30 0 30 5]{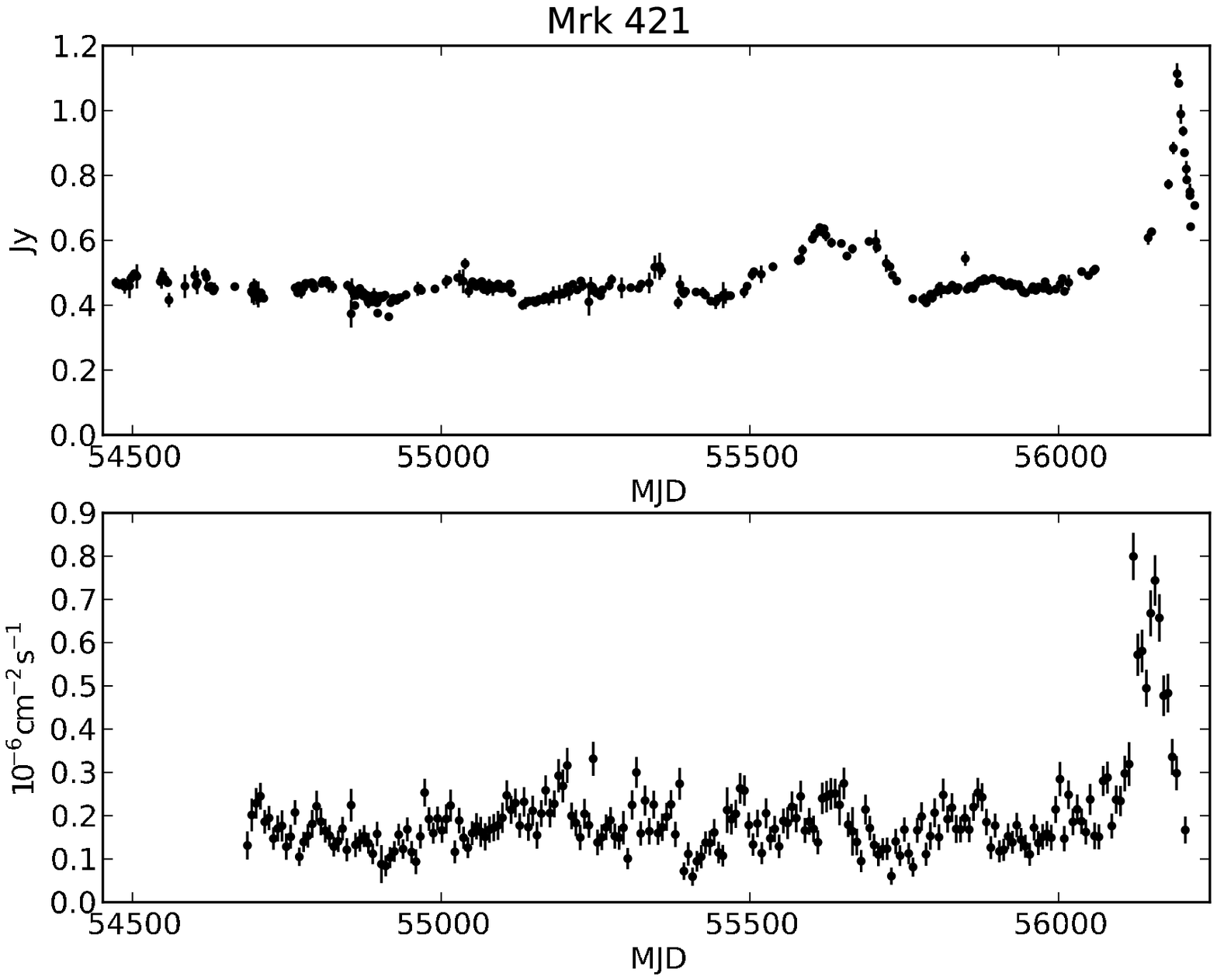}
\includegraphics[angle=0,width=8.5cm, trim=30 0 30 5]{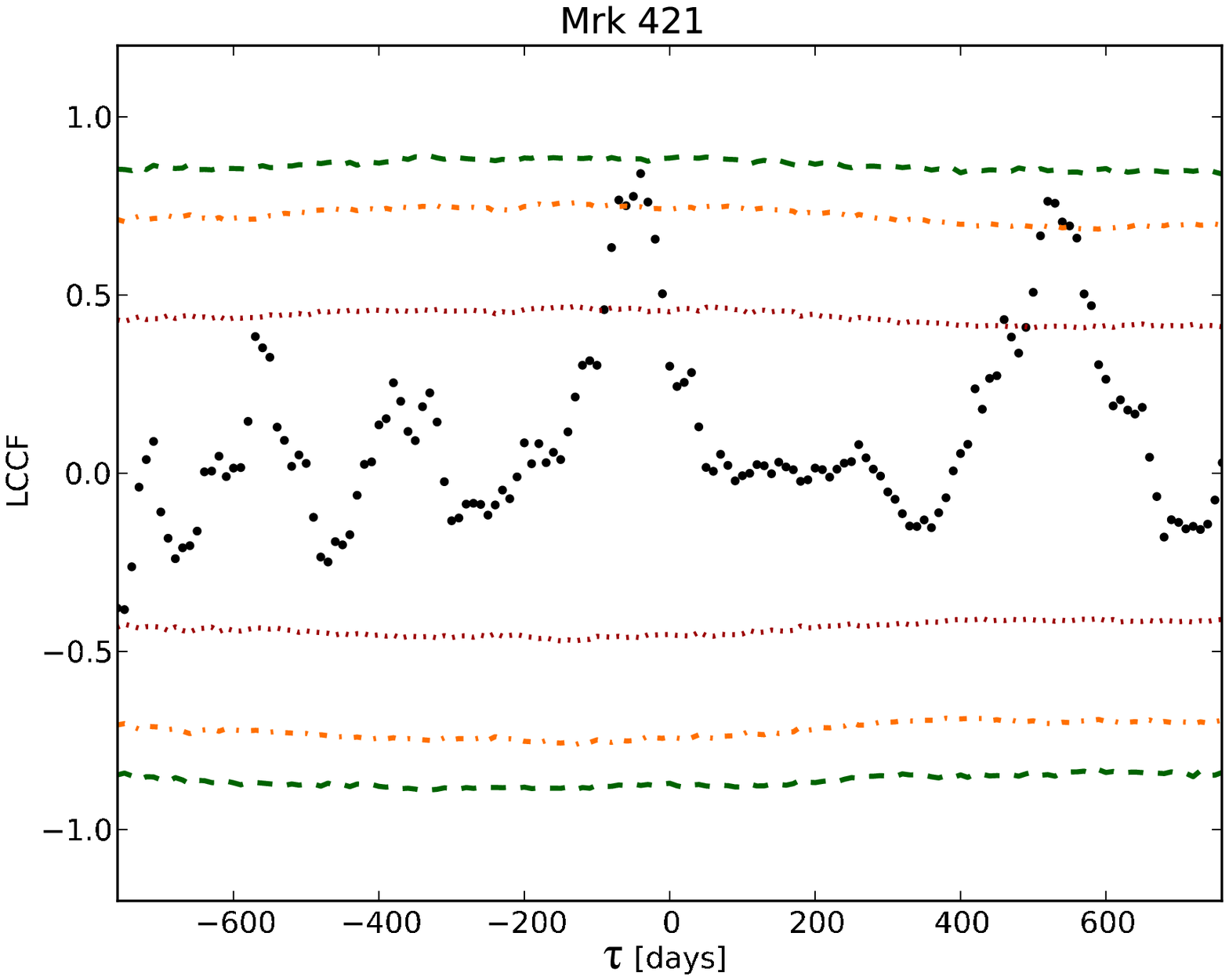}
\caption{Light-curves (left) and cross-correlation (right) for Mrk\,421. The most significant peak is at $-40\pm9\,$d with 98.96\% significance. Colors and line styles as in Figure~\ref{xcorr_combined}.}
\label{xcorr_mrk421}
\end{figure*}

Mrk\,421 does not show significant correlated variability when analysed as part of the uniform sample described in Section \ref{observations}. Furthermore, neither the radio nor the gamma-ray PSD can be fit so the population averages were used as described in Section \ref{time-lags-and-significance}. To include the radio and gamma-ray flares we extended the light-curves beyond the period used for the uniform sample (Section \ref{observations}). We repeated the analysis using these extended light-curves and found $0.6<\beta_{\rm radio}<2.0$, with best fit value 1.8, and $1.6<\beta_{\gamma}<2.1$, with best fit value 1.6. The cross-correlation peak at $-40\pm9\,$d has a significance between 96.16\% and 99.99\% depending on the PSD model. The significance obtained using the best-fit PSD models is 98.96\% (Figure~\ref{xcorr_mrk421}). This result should be treated with caution: extending the data set after noticing the flare is ``a posteriori" statistics, and as such cannot be used to make inferences about the rate at which significant correlations are found in the general blazar population.

\section{Interpretation of the time delays}

\begin{figure*}
\begin{center}
\includegraphics[width=12.0cm, angle=0, trim=0 0 0 0]{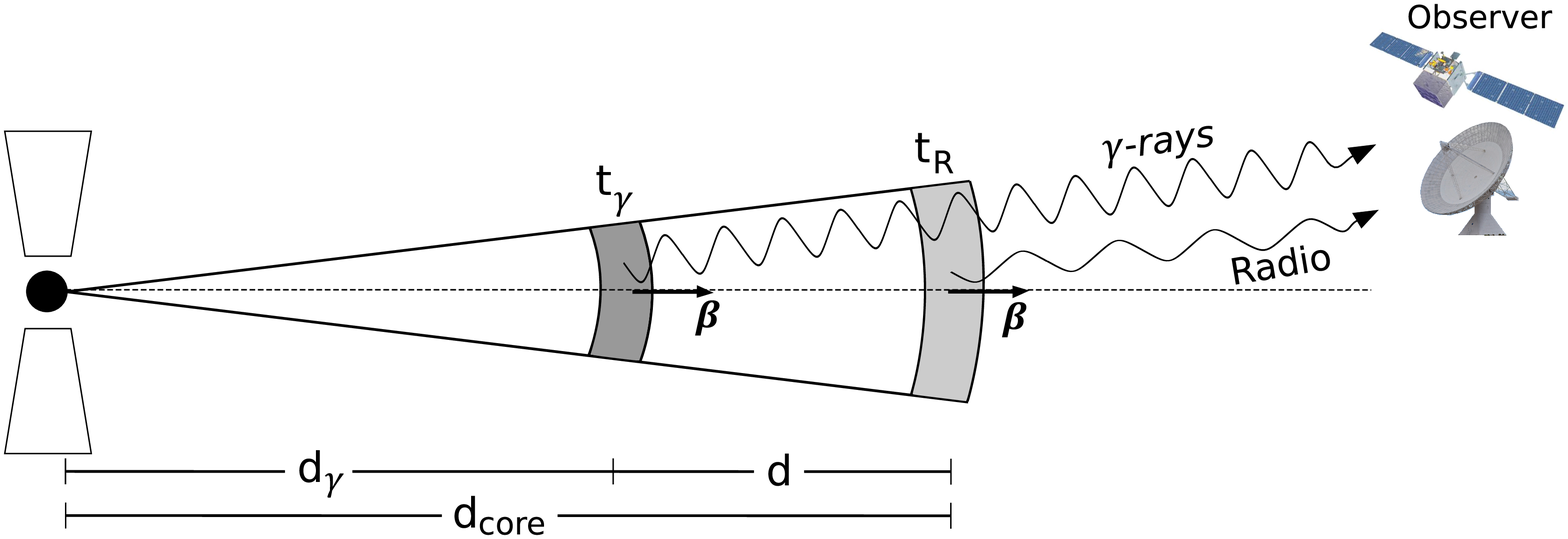}
\caption{Model for the interpretation of time lags. The central engine launches a jet in which disturbances propagate at speed $\beta c$. A moving disturbance (shaded area) is depicted at two times: $t_{\gamma}$ at which gamma-ray emission peaks and $t_{\rm R}$ for the peak of radio emission when crossing the radio core.}
\label{delay_model_schematic}
\end{center}
\end{figure*}

The duration of the correlated events is typically a few hundred days and a detailed model is needed to understand the relationship between the lags and the location of the emission regions. Here, we ignore the flare duration and tentatively interpret the delays using a model in which a moving emission region, confined to the jet, produces the radio and gamma-ray activity. This region moves outward at the bulk jet speed $\beta c$ (Figure~\ref{delay_model_schematic}), and corresponds to the moving disturbances observed with very long baseline interferometry (VLBI). The gamma-ray flare becomes observable at distance $d_{\gamma}$ from the central engine, after crossing the surface of unit gamma-ray opacity \citep[gamma-sphere,][]{blandford_and_levinson_1995}. Likewise, the radio flare becomes observable upon crossing the surface of unit radio opacity (``radio core''), at distance $d_{\rm core}$ from the central engine \citep[][]{blandford_1979}.

The time lag between these wavebands provides an estimate of the interval between the emergence of gamma-ray and radio radiation. The distance travelled by the emission region between the peaks in gamma-ray and radio emission is

\begin{equation}
d = \frac{\Gamma D\,\beta c\,\Delta t}{(1+z)},
\end{equation}
where $\Gamma$ is the bulk Lorentz factor, $D$ is the Doppler factor, $\Delta t$ is the time lag and $z$ is the redshift \citep[][]{pushkarev_2010}. The apparent jet speed, $\beta_{\rm app}$, is determined from VLBI monitoring and the Doppler factor is estimated from the radio variability time scale \citep[][]{hovatta+2009}. Doppler factors from this method have a typical 27\% scatter for individual flares in a given source, which we adopt as the uncertainty in $D$. From $D$ and $\beta_{\rm app}$, we obtain $\Gamma$ and the jet viewing angle $\theta$ \citep[e.g.,][]{hovatta+2009}. $\beta_{\rm app}$ and $D$ are not measured simultaneously with our observations; we assume them constant in our calculations.

We estimate $d_{\gamma}=d_{\rm core}-d$, where $d_{\rm core}$ is determined from VLBI measurements of the angular diameter of the radio core, $\theta_{\rm core}$. This, plus the intrinsic opening angle, $\alpha_{\rm int}$, and redshift, gives 
\begin{equation}
\label{d_core}
d_{\rm core}\sim\frac{(\theta_{\rm core}/2)d_{\rm A}}{\tan(\alpha_{\rm int}/2)},
\end{equation}
where $d_{\rm A}$ is the angular diameter distance, obtained assuming a $\Lambda$CDM cosmology with $H_0=70$ km s$^{-1}$ Mpc$^{-1}$, $\Omega_{\rm m}=0.27$ and $\Omega_{\Lambda}=0.73$ \citep[][]{komatsu+2011}. Equation~\ref{d_core} is only valid for a conical jet with vertex at the central engine. However, there is observational evidence for collimation in the M87 jet, which we use as a prototype for the collimation properties of other sources where no such information is available. \citet[][]{asada_nakamura_2012} model the jet profile as $z_{\rm jet}\propto~r^a$,  where  $r$ is the radius of the jet cross-section at distance $z_{\rm jet}$ from the central engine, and found $a=1.73\pm0.05$ for $z_{\rm jet}\la2.5\times10^5\,r_{\rm s}$, where $r_{\rm s}$ is the Schwarzschild radius, and $a=0.96\pm0.1$  at distances outside the collimation zone. Assuming the radio core is in the collimation zone and setting $r$ and  $dr/dz_{\rm jet}$ equal for both models

\begin{equation}
\label{coll_cone_correction}
d_{\rm core}({\rm coll})=\frac{1}{a}d_{\rm core}({\rm cone}).
\end{equation}

This model reduces our estimate of $d_{\rm core}$ by a factor of 1.73. We thus obtain lower and upper limits on $d_{\rm core}$ using these alternatives.

Detailed distance estimates are provided below for AO\,0235+164 the highest significance case and Mrk\,421 which has the sharpest cross-correlation peak. For PKS\,1502+106 only the final result is given as a reference, and for B2\,2308+34 there are no published VLBI results which makes it impossible to provide a constraint. A summary of the results for AO\,0235+164 is given in Table \ref{d_gamma_table_summary}.

\begin{table*}
\begin{minipage}{126mm}
\caption{Results of the distance estimates for the different jet components in the most significant case.$^{(a)}$}
\label{d_gamma_table_summary}
\begin{tabular}{l c c c c c}
\hline
Source	& $d$ & $d_{\rm core}({\rm coll})$ & $d_{\rm core}({\rm cone})$ & $d_{\gamma}({\rm coll})$ & $d_{\gamma}({\rm cone})$ \\
	& [pc]	& [pc] & [pc] & [pc] & [pc] \\
\hline
AO\,0235+164, $\tau = -150 \pm 8\,$d   	& $37 \pm 23$	& $\ga 23 \pm 6$	& $\ga 40 \pm 11$	& $\ga -14 \pm 24$$^{(b)}$	& $\ga 3 \pm 25$	\\
AO\,0235+164, $\tau = -30 \pm 9\,$d   	& $8 \pm 5$	& $\ga 23 \pm 6$	& $\ga 40 \pm 11$	& $\ga 15 \pm 8$	& $\ga 32 \pm 12$\\
\hline
\end{tabular}
\medskip
(a): Columns: $d$: distance travelled by the emission region between the peaks in gamma-ray and radio emission. $d_{\rm core}{\rm (coll/cone)}$: distance between radio core and central engine with and without collimation. $d_{\gamma}{\rm (coll/cone)}$: Location of the peak of the gamma-ray emission with respect to the central engine.

(b): The negative value is an artifact produced by the large measurement errors.
\end{minipage}
\end{table*}

\paragraph*{Estimation of $d$:}
\label{d_estimate_section}

For AO\,0235+164 we have $D=24$ \citep[][]{hovatta+2009} but no $\beta_{\rm app}$ since its jet is unresolved in 15\,GHz VLBI \citep[][]{lister+2009}. We assume the source is seen at the critical angle, $\theta_{\rm cr}=\theta=2.4\degr$. We obtain $d=37.3\pm22.8$ pc for $\tau=-150\pm8\,$d, the most significant time lag, and $d=7.5\pm5.3$ pc for the peak at $\tau=-30\pm9\,$d. For comparison, if we use $\theta=\theta_{\rm cr}/2$, we obtain $d=20\pm15$~pc ($3.9\pm3.2$~pc) for the peak at $-150\,$d ($-30\,$d). If $\theta=0$, we obtain $d=18\pm12$~pc ($3.7\pm2.6$~pc).

For Mrk\,421, we use a preliminary variability Doppler factor for the recent flare of $D=4$ \citep[][]{richards+2013}. $\beta_{\rm app}$ is uncertain, with jet components consistent with being stationary \citep[][]{lico+2012}. Assuming $\theta\sim4\degr$ (\citealt{lico+2012} estimate 2\degr--5\degr), then $\Gamma\sim2.2$ and $d\sim0.2$~pc. It is difficult to estimate uncertainties because of the limited knowledge of the jet properties.

\paragraph*{Estimation of $d_{\rm core}$:}

The core angular size (FWHM) have been measured for AO\,0235+164 ($\theta_{\rm core}=0.21\pm0.06$ mas) \citep[][]{lister+2009}. Here, we have averaged multiple epochs, with uncertainties estimated from their scatter. For Mrk\,421 we use $\theta_{\rm core}=0.16$ mas \citep[][]{kovalev+2005}, assuming an error of $0.05$ mas, the angular resolution of the observations.

For the intrinsic opening angle we use $\alpha_{\rm int}\la2.4\degr$ for AO\,0235+164, which is the critical angle upper limit from Section \ref{d_estimate_section}, consistent with what is used by \citet[][]{agudo_ao_2011}. For Mrk\,421, we adopt $\alpha_{\rm int}=2.4\degr$, the mean value for BL~Lacs from \citet[][]{pushkarev+2009}. 

The estimates of $d_{\rm core}$ for a conical jet are $\ga40\pm11$ pc for AO\,0235+164, and about 2.4\,pc for Mrk\,421. For a collimated jet we obtain $d_{\rm core}\ga23\pm6$ pc for AO\,0235+164, and about 1.4 pc for Mrk\,421.

Similar estimates can be made for PKS\,1502+106, resulting in $d_{\gamma}$ of $22\pm15$ pc for a conical jet and $12\pm9$ pc for the collimated jet case.

\section{Conclusions}

Out of 41 sources for which a detailed correlation analysis is possible, three show correlations with larger than 2.25$\sigma$ significance, with only one of those larger than 3$\sigma$. In all cases radio variations lag behind gamma-ray variations, suggesting that the gamma-ray emission originates upstream of the radio emission. We use a simple model to tentatively estimate the distance from the black hole at which the gamma-ray emission is produced. Due to correlation peak breadth and uncertain jet parameters, these estimates have large uncertainties. In particular AO\,0235+164 shows two peaks in its cross-correlation with comparable amplitude and equivalent significance, leading to a highly uncertain location for the gamma-ray emission site.

These results show that correlations between radio and gamma-ray light-curves of blazars are only found in a minority of the sources over a four-year period. This could indicate a complex multiwavelength connection not detectable with the tools and data we use. A better understanding of this connection requires continuation of the OVRO and \emph{Fermi} monitoring and will benefit from the addition of polarization and other wavebands and methods that provide additional information.

\section*{Acknowledgments}
We thank Russ Keeney for his support at OVRO. The OVRO program is supported in part by NASA grants NNX08AW31G and NNX11A043G and NSF grants AST-0808050 and AST-1109911. T.H. was supported by the Jenny and Antti Wihuri foundation and Academy of Finland project number 267324. Support from MPIfR for upgrading the OVRO 40-m telescope receiver is acknowledged. W.M. thanks Jeffrey Scargle, James Chiang, Stefan Larsson and Iossif Papadakis for discussions. The National Radio Astronomy Observatory is a facility of the National Science Foundation operated under cooperative agreement by Associated Universities, Inc. The $Fermi$ LAT Collaboration acknowledges support from a number of agencies and institutes for both development and the operation of the LAT as well as scientific data analysis. These include NASA and DOE in the United States, CEA/Irfu and IN2P3/CNRS in France, ASI and INFN in Italy, MEXT, KEK, and JAXA in Japan, and the K.~A.~Wallenberg Foundation, the Swedish Research Council and the National Space Board in Sweden. Additional support from INAF in Italy and CNES in France for science analysis during the operations phase is also gratefully acknowledged. We thank the anonymous referee for constructive comments that greatly improved the presentation of some sections of this paper.


\bsp 
\label{lastpage}


\newgeometry{left=0.5in, right=0.5in, bottom=0.8in, top=0.8in}

\setcounter{table}{0}


\begin{table*}
\caption{Cross-correlation significance results \label{xcorr_sig_table}}
\scalebox{0.72}{
\begin{tabular}{l c c c c c c c c c c r@{$\,\,\pm\,\,$}l c c c c c c l}
\hline
Source & Name & Class & Class & $z$ & $\beta_{\rm radio}^{\rm best}$ & $\beta_{\rm radio}^{\rm low}$ & $\beta_{\rm radio}^{\rm up}$ & $\beta_{\gamma}^{\rm best}$ & $\beta_{\gamma}^{\rm low}$ & $\beta_{\gamma}^{\rm up}$ & \multicolumn{2}{c}{$\tau$$^{©}$} & DCF & Sig. & Sig.$_{\rm low}$ & Sig.$_{\rm up}$ & Sig.$_{\rm unc}$ & Sig.$_{\sigma}$ & Flags \\
name & 2FGL & optical$^{(a)}$ & SED$^{(a)}$ & $^{(a)}$ & $^{(b)}$ & $^{(b)}$ & $^{(b)}$ & $^{(b)}$ & $^{(b)}$ & $^{(b)}$ & \multicolumn{2}{c}{[d]} & $^{(c)}$ & \,\,\%\,$^{(c)}$ & \,\,\%\,$^{(c)}$ & \,\,\%\,$^{(c)}$ & \,\,\%\,$^{(c)}$ & \,\,\,$\sigma$\,$^{(c)}$ & $^{(d)}$ \\
\hline
4C +01.02 & J0108.6+0135 & FSRQ & LSP & 2.099 & 2.3 & \ldots & \ldots & 1.6 & \ldots & \ldots & $-$340 & 16 & 0.33 & 58.64 & \ldots & \ldots & 0.49 & 0.82 & ng                                         \\
S2 0109+22 & J0112.1+2245 & BL Lac & ISP & 0.265 & 2.0 & 1.4 & 2.4 & 0.9 & 0.0 & 1.8 & $-$380 & 13 & 0.24 & 59.63 & 36.05 & 94.45 & 0.48 & 0.84 & \ldots                                             \\
4C 31.03 & J0112.8+3208 & FSRQ & LSP & 0.603 & 2.3 & \ldots & \ldots & 1.6 & \ldots & \ldots & 190 & 12 & 0.36 & 54.65 & \ldots & \ldots & 0.5 & 0.75 & tg                                              \\
OC 457 & J0136.9+4751 & FSRQ & LSP & 0.859 & 1.6 & 1.4 & 1.9 & 1.6 & \ldots & \ldots & $-$230 & 14 & 0.62 & 95.75 & 92.92 & 97.76 & 0.19 & 2.03 & \ldots                                                \\
PKS 0215+015 & J0217.9+0143 & FSRQ & LSP & 1.721 & 2.3 & \ldots & \ldots & 1.6 & \ldots & \ldots & $-$60 & 15 & 0.38 & 65.75 & \ldots & \ldots & 0.46 & 0.95 & \ldots                                   \\
S4 0218+35 & J0221.0+3555 & FSRQ & \ldots & 0.944 & 2.3 & \ldots & \ldots & 1.6 & \ldots & \ldots & 190 & 12 & 0.51 & 93.08 & \ldots & \ldots & 0.25 & 1.82 & ng                                    \\
3C 66A & J0222.6+4302 & BL Lac & ISP & \ldots & 1.9 & 0.4 & 2.5 & 0.6 & 0.2 & 1.2 & 460 & 14 & 0.27 & 81.16 & 60.78 & 99.96 & 0.39 & 1.32 & \ldots                                                      \\
4C +28.07 & J0237.8+2846 & FSRQ & LSP & 1.206 & 2.7 & 2.5 & 3.0 & 1.6 & \ldots & \ldots & $-$140 & 13 & 0.63 & 83.59 & 81.78 & 85.31 & 0.37 & 1.39 & \ldots                                             \\
\textbf{AO 0235+164} & \textbf{J0238.7+1637} & \textbf{BL Lac} & \textbf{LSP} & \textbf{0.94} & \textbf{2.3} & \textbf{\ldots} & \textbf{\ldots} & \textbf{0.1} & \textbf{0.0} & \textbf{1.0} & \textbf{$-$150} & \textbf{8} & \textbf{0.91} & \textbf{99.99} & \textbf{99.99} & \textbf{99.99} & \textbf{\ldots} & \textbf{3.89} & \textbf{\ldots}                                          \\
NGC 1275 & J0319.8+4130 & Radio Gal & \ldots & 0.018 & 2.3 & \ldots & \ldots & 1.6 & 1.1 & 2.2 & $-$420 & 13 & 0.59 & 81.54 & 73.56 & 93.0 & 0.39 & 1.33 & \ldots                                       \\
PKS 0420$-$01 & J0423.2$-$0120 & FSRQ & LSP & 0.916 & 2.5 & 2.2 & 2.8 & 1.6 & \ldots & \ldots & $-$20 & 16 & 0.49 & 76.65 & 75.2 & 79.57 & 0.42 & 1.19 & \ldots                                         \\
PKS 0440$-$00 & J0442.7$-$0017 & FSRQ & LSP & 0.844 & 2.3 & \ldots & \ldots & 0.7 & 0.1 & 2.3 & 420 & 32 & 0.22 & 59.27 & 31.11 & 78.29 & 0.47 & 0.83 & \ldots                                          \\
TXS 0506+056 & J0509.4+0542 & BL Lac & ISP & 0.0 & 2.2 & 0.6 & 2.7 & 1.6 & \ldots & \ldots & 450 & 15 & 0.25 & 61.54 & 58.89 & 96.86 & 0.5 & 0.87 & tg, ng                                              \\
B2 0716+33 & J0719.3+3306 & FSRQ & LSP & 0.779 & 2.3 & \ldots & \ldots & 1.6 & \ldots & \ldots & 140 & 8 & 0.61 & 90.05 & \ldots & \ldots & 0.31 & 1.65 & \ldots                                        \\
S5 0716+71 & J0721.9+7120 & BL Lac & ISP & 0.0 & 2.3 & \ldots & \ldots & 1.9 & 1.4 & 2.3 & $-$200 & 11 & 0.37 & 44.89 & 39.86 & 55.97 & 0.49 & 0.6 & \ldots                                             \\
4C +14.23 & J0725.3+1426 & FSRQ & LSP & 1.038 & 2.3 & \ldots & \ldots & 0.5 & 0.1 & 1.0 & 150 & 13 & 0.19 & 61.5 & 43.28 & 76.51 & 0.48 & 0.87 & \ldots                                                 \\
PKS 0736+01 & J0739.2+0138 & FSRQ & LSP & 0.189 & 2.3 & \ldots & \ldots & 1.6 & \ldots & \ldots & $-$360 & 15 & 0.5 & 79.7 & \ldots & \ldots & 0.39 & 1.27 & ng                                         \\
GB6 J0742+5444 & J0742.6+5442 & FSRQ & LSP & 0.723 & 1.9 & 0.6 & 2.9 & 1.6 & \ldots & \ldots & $-$190 & 9 & 0.69 & 92.09 & 83.89 & 99.99 & 0.27 & 1.76 & \ldots                                         \\
PKS 0805$-$07 & J0808.2$-$0750 & FSRQ & LSP & 1.837 & 2.0 & 1.6 & 2.5 & 0.5 & 0.1 & 1.1 & $-$150 & 16 & 0.59 & 99.52 & 88.86 & 99.99 & 0.07 & 2.82 & \ldots                                             \\
PKS 0829+046 & J0831.9+0429 & BL Lac & LSP & 0.174 & 1.9 & 0.6 & 2.3 & 1.6 & \ldots & \ldots & 150 & 16 & 0.42 & 81.09 & 75.95 & 99.89 & 0.4 & 1.31 & ng                                            \\
4C +71.07 & J0841.6+7052 & FSRQ & LSP & 2.218 & 2.3 & \ldots & \ldots & 1.6 & \ldots & \ldots & 210 & 11 & 0.63 & 91.74 & \ldots & \ldots & 0.28 & 1.74 & tg, ng                                        \\
OJ 287 & J0854.8+2005 & BL Lac & ISP & 0.306 & 2.3 & \ldots & \ldots & 1.6 & \ldots & \ldots & $-$70 & 16 & 0.38 & 62.48 & \ldots & \ldots & 0.48 & 0.89 & ng                                           \\
PKS 0906+01 & J0909.1+0121 & FSRQ & LSP & 1.026 & 2.3 & \ldots & \ldots & 1.6 & \ldots & \ldots & 510 & 16 & 0.39 & 68.85 & \ldots & \ldots & 0.48 & 1.01 & \ldots                                          \\
S4 0917+44 & J0920.9+4441 & FSRQ & LSP & 2.189 & 2.3 & \ldots & \ldots & 1.6 & 0.8 & 2.1 & $-$460 & 12 & 0.49 & 70.51 & 62.12 & 92.86 & 0.48 & 1.05 & \ldots                                                \\
MG2 J101241+2439 & J1012.6+2440 & FSRQ & \ldots & 1.805 & 2.3 & \ldots & \ldots & 1.6 & \ldots & \ldots & 490 & 49 & 0.57 & 99.51 & \ldots & \ldots & 0.07 & 2.81 & tr, tg, nr, ng                      \\
4C +01.28 & J1058.4+0133 & BL Lac & LSP & 0.888 & 2.3 & \ldots & \ldots & 1.6 & \ldots & \ldots & 510 & 15 & 0.6 & 93.42 & \ldots & \ldots & 0.25 & 1.84 & ng                                           \\
Mrk 421 & J1104.4+3812 & BL Lac & HSP & 0.031 & 2.3 & \ldots & \ldots & 1.6 & \ldots & \ldots & $-$500 & 10 & 0.39 & 73.78 & \ldots & \ldots & 0.43 & 1.12 & ng                                         \\
PKS 1124$-$186 & J1126.6$-$1856 & FSRQ & LSP & 1.048 & 2.0 & 1.6 & 2.4 & 1.6 & \ldots & \ldots & 10 & 11 & 0.76 & 97.62 & 95.88 & 99.2 & 0.15 & 2.26 & \ldots                                           \\
Ton 599 & J1159.5+2914 & FSRQ & LSP & 0.725 & 2.1 & 1.8 & 2.6 & 1.0 & 0.5 & 1.6 & $-$70 & 18 & 0.42 & 79.05 & 55.61 & 95.39 & 0.39 & 1.25 & \ldots                                                      \\
1ES 1215+303 & J1217.8+3006 & BL Lac & HSP & 0.13 & 2.3 & \ldots & \ldots & 1.6 & \ldots & \ldots & 120 & 9 & 0.5 & 91.88 & \ldots & \ldots & 0.27 & 1.74 & ng                                          \\
4C +21.35 & J1224.9+2122 & FSRQ & LSP & 0.434 & 2.4 & 0.9 & 2.7 & 0.4 & 0.2 & 0.8 & $-$380 & 10 & 0.59 & 99.78 & 96.51 & 99.99 & 0.05 & 3.06 & \ldots                                                   \\
3C 273 & J1229.1+0202 & FSRQ & LSP & 0.158 & 2.2 & 0.6 & 2.8 & 0.8 & 0.4 & 1.1 & $-$240 & 16 & 0.41 & 86.32 & 68.08 & 99.99 & 0.33 & 1.49 & \ldots                                                      \\
MG1 J123931+0443 & J1239.5+0443 & FSRQ & LSP & 1.761 & 2.3 & \ldots & \ldots & 1.6 & \ldots & \ldots & $-$50 & 15 & 0.67 & 89.01 & \ldots & \ldots & 0.3 & 1.6 & \ldots                                 \\
3C 279 & J1256.1$-$0547 & FSRQ & LSP & 0.536 & 2.4 & 2.0 & 2.7 & 1.6 & 1.2 & 2.0 & 190 & 10 & 0.47 & 65.03 & 54.86 & 80.94 & 0.48 & 0.94 & \ldots                                                       \\
OP 313 & J1310.6+3222 & FSRQ & LSP & 0.997 & 2.2 & 1.9 & 2.4 & 1.6 & \ldots & \ldots & 500 & 74 & 0.32 & 49.79 & 47.75 & 54.35 & 0.5 & 0.67 & \ldots                                                    \\
GB 1310+487 & J1312.8+4828 & FSRQ & LSP & 0.501 & 2.3 & \ldots & \ldots & 0.3 & 0.0 & 1.0 & $-$350 & 15 & 0.36 & 93.09 & 77.42 & 96.31 & 0.25 & 1.82 & \ldots                                        \\
PKS 1329$-$049 & J1332.0$-$0508 & FSRQ & LSP & 2.15 & 2.2 & 1.4 & 2.9 & 0.3 & 0.1 & 0.8 & $-$90 & 15 & 0.51 & 99.56 & 90.36 & 99.98 & 0.07 & 2.85 & \ldots                                              \\
B3 1343+451 & J1345.4+4453 & FSRQ & LSP & 2.534 & 2.1 & 0.6 & 2.6 & 1.6 & \ldots & \ldots & 30 & 13 & 0.51 & 67.5 & 62.53 & 99.88 & 0.48 & 0.98 & \ldots                                                \\
PKS 1424+240 & J1427.0+2347 & BL Lac & HSP & 0.0 & 2.3 & \ldots & \ldots & 1.6 & \ldots & \ldots & 110 & 13 & 0.45 & 95.59 & \ldots & \ldots & 0.2 & 2.01 & tr, tg, nr, ng                          \\
\textbf{PKS 1502+106} & \textbf{J1504.3+1029} & \textbf{FSRQ} & \textbf{LSP} & \textbf{1.839} & \textbf{2.5} & \textbf{2.2} & \textbf{2.8} & \textbf{1.6} & \textbf{\ldots} & \textbf{\ldots} & \textbf{$-$40} & \textbf{13} & \textbf{0.87} & \textbf{98.09} & \textbf{97.54} & \textbf{98.7} & \textbf{0.13} & \textbf{2.34} & \textbf{\ldots}                                            \\
PKS 1510$-$08 & J1512.8$-$0906 & FSRQ & LSP & 0.36 & 2.3 & 1.6 & 2.9 & 1.6 & \ldots & \ldots & $-$60 & 6 & 0.65 & 83.59 & 78.01 & 92.58 & 0.37 & 1.39 & \ldots                                          \\
B2 1520+31 & J1522.1+3144 & FSRQ & LSP & 1.484 & 2.3 & \ldots & \ldots & 0.7 & 0.4 & 1.0 & 350 & 8 & 0.45 & 95.65 & 90.11 & 99.0 & 0.2 & 2.02 & \ldots                                                  \\
GB6 J1542+6129 & J1542.9+6129 & BL Lac & ISP & 0.0 & 2.3 & \ldots & \ldots & 1.6 & \ldots & \ldots & 360 & 16 & 0.38 & 78.75 & \ldots & \ldots & 0.41 & 1.25 & tg, ng                                   \\
PG 1553+113 & J1555.7+1111 & BL Lac & HSP & 0.0 & 2.3 & \ldots & \ldots & 1.6 & \ldots & \ldots & 530 & 17 & 0.43 & 99.69 & \ldots & \ldots & 0.06 & 2.96 & tr, tg, nr, ng                              \\
4C +38.41 & J1635.2+3810 & FSRQ & LSP & 1.813 & 2.1 & 1.4 & 2.9 & 1.5 & 1.1 & 1.8 & 500 & 8 & 0.79 & 96.28 & 90.24 & 99.99 & 0.2 & 2.08 & \ldots                                                        \\
Mrk 501 & J1653.9+3945 & BL Lac & HSP & 0.034 & 2.3 & \ldots & \ldots & 1.6 & \ldots & \ldots & $-$480 & 12 & 0.5 & 98.11 & \ldots & \ldots & 0.13 & 2.35 & tg, ng                                      \\
B3 1708+433 & J1709.7+4319 & FSRQ & LSP & 1.027 & 2.3 & \ldots & \ldots & 1.6 & \ldots & \ldots & $-$50 & 12 & 0.59 & 80.86 & \ldots & \ldots & 0.39 & 1.31 & \ldots                                    \\
PKS 1730$-$13 & J1733.1$-$1307 & FSRQ & LSP & 0.902 & 2.0 & 1.5 & 2.4 & 1.6 & \ldots & \ldots & $-$260 & 14 & 0.5 & 73.67 & 68.25 & 83.55 & 0.44 & 1.12 & tg                                        \\
S4 1749+70 & J1748.8+7006 & BL Lac & ISP & 0.77 & 2.2 & 1.4 & 2.7 & 0.4 & 0.0 & 1.1 & 230 & 10 & 0.55 & 99.48 & 87.86 & 99.99 & 0.07 & 2.79 & \ldots                                                 \\
S5 1803+784 & J1800.5+7829 & BL Lac & LSP & 0.68 & 2.3 & \ldots & \ldots & 0.4 & 0.0 & 0.9 & $-$430 & 11 & 0.47 & 98.43 & 89.17 & 99.89 & 0.12 & 2.42 & tg                                           \\
4C +56.27 & J1824.0+5650 & BL Lac & LSP & 0.664 & 1.9 & 0.4 & 2.9 & 1.6 & \ldots & \ldots & $-$240 & 11 & 0.46 & 80.36 & 74.26 & 99.99 & 0.39 & 1.29 & tg, ng                                           \\
B2 1846+32A & J1848.5+3216 & FSRQ & LSP & 0.798 & 2.2 & 1.9 & 2.7 & 1.6 & \ldots & \ldots & $-$300 & 12 & 0.53 & 65.95 & 61.37 & 69.83 & 0.47 & 0.95 & \ldots                                           \\
S4 1849+67 & J1849.4+6706 & FSRQ & LSP & 0.657 & 1.9 & 0.8 & 2.5 & 0.6 & 0.2 & 1.2 & $-$40 & 10 & 0.38 & 90.61 & 61.42 & 99.91 & 0.3 & 1.68 & \ldots                                                    \\
1ES 1959+650 & J2000.0+6509 & BL Lac & HSP & 0.047 & 2.3 & \ldots & \ldots & 1.6 & \ldots & \ldots & $-$80 & 13 & 0.41 & 90.97 & \ldots & \ldots & 0.3 & 1.69 & tg, ng                                  \\
PKS 2023$-$07 & J2025.6$-$0736 & FSRQ & LSP & 1.388 & 2.3 & \ldots & \ldots & 1.6 & \ldots & \ldots & 130 & 12 & 0.55 & 72.89 & \ldots & \ldots & 0.44 & 1.1 & \ldots                                   \\
OX 169 & J2143.5+1743 & FSRQ & LSP & 0.211 & 2.3 & \ldots & \ldots & 0.0 & 0.0 & 0.5 & $-$320 & 11 & 0.34 & 99.04 & 89.26 & 98.96 & 0.1 & 2.59 & \ldots                                           \\
BL Lacertae & J2202.8+4216 & BL Lac & ISP & 0.069 & 2.1 & 0.9 & 2.7 & 2.0 & 1.5 & 2.4 & $-$160 & 14 & 0.71 & 85.27 & 72.37 & 99.99 & 0.36 & 1.45 & \ldots                                               \\
PKS 2201+171 & J2203.4+1726 & FSRQ & LSP & 1.076 & 2.0 & 1.5 & 2.3 & 1.6 & \ldots & \ldots & 530 & 10 & 0.58 & 93.29 & 92.2 & 97.46 & 0.26 & 1.83 & ng                                                  \\
PKS 2227$-$08 & J2229.7$-$0832 & FSRQ & LSP & 1.56 & 2.8 & 2.4 & 3.1 & 1.6 & \ldots & \ldots & 310 & 14 & 0.51 & 80.95 & 79.82 & 83.36 & 0.41 & 1.31 & ng                                               \\
CTA 102 & J2232.4+1143 & FSRQ & LSP & 1.037 & 2.4 & 1.7 & 2.8 & 1.6 & \ldots & \ldots & $-$430 & 8 & 0.42 & 55.1 & 51.02 & 66.17 & 0.5 & 0.76 & \ldots                                                  \\
B2 2234+28A & J2236.4+2828 & BL Lac & LSP & 0.795 & 1.9 & 0.6 & 2.3 & 1.6 & \ldots & \ldots & 110 & 14 & 0.35 & 63.42 & 58.81 & 99.39 & 0.48 & 0.9 & \ldots                                             \\
3C 454.3 & J2253.9+1609 & FSRQ & LSP & 0.859 & 2.4 & 1.9 & 2.6 & 1.6 & \ldots & \ldots & $-$80 & 18 & 0.55 & 71.96 & 70.12 & 78.46 & 0.46 & 1.08 & \ldots                                               \\
\textbf{B2 2308+34} & \textbf{J2311.0+3425} & \textbf{FSRQ} & \textbf{LSP} & \textbf{1.817} & \textbf{2.1} & \textbf{0.6} & \textbf{2.7} & \textbf{0.2} & \textbf{0.0} & \textbf{0.9} & \textbf{$-$120} & \textbf{14} & \textbf{0.73} & \textbf{99.99} & \textbf{99.33} & \textbf{99.99} & \textbf{\ldots} & \textbf{3.89} & \textbf{\ldots}                                             \\
\hline
\end{tabular}
}
\medskip
\\
(a): Optical class, SED class and redshifts from Ackermann et al. (2011b). $z=0.0$ indicate that redshift could not be evaluated with available optical spectrum. (b): $\beta_{\rm waveband}^{\rm best/low/up}$: PSD power-law index for given  ``waveband'': ``best'' for best fit, ``low'' for lower limit and ``up'' for upper limit. $\tau$: radio/gamma-ray time lag, negative values indicate radio lags gamma-ray variations. DCF: discrete correlation function estimate. Sig.: Significance of the correlation. Sig$_{\rm low/up/unc/\sigma}$: Significance lower limit, upper limit, uncertainty and significance in units of standard deviations. (c): $\tau$: radio/gamma-ray time lag, negative values indicate radio lags gamma-ray variations. DCF: discrete correlation function estimate. Sig.: Significance of the correlation. Sig$_{\rm low/up/unc/\sigma}$: Significance lower limit, upper limit, uncertainty and significance in units of standard deviations. (d): Flags are: noisy light curves in radio (nr) and gamma-ray band (ng); trends in radio (tr) and gamma-ray band (tg) (see text).
\end{table*}

\end{document}